\documentclass{article}
\usepackage{ijcai20}

\usepackage{times}

\usepackage{soul}
\usepackage{url}
\usepackage[hidelinks]{hyperref}
\usepackage[utf8]{inputenc}
\usepackage[small]{caption}
\usepackage{graphicx}
\usepackage{amsmath}
\usepackage{booktabs}
\usepackage{balance}
\usepackage{siunitx}
\usepackage[linesnumbered,ruled,vlined,procnumbered]{algorithm2e}
\usepackage{diagbox}
\usepackage{comment}
\usepackage{tikz}
\usetikzlibrary{calc}
\usetikzlibrary{backgrounds,positioning}
\usetikzlibrary{decorations.pathreplacing}
\usepackage{subcaption}
\usepackage{diagbox}
\usepackage{comment}
\usepackage{xcolor}
\usepackage{etoolbox}

\makeatletter
\AtBeginEnvironment{procedure}{\let\c@algocf\c@procedure}
\makeatother

\usepackage{amsmath}
\DeclareMathOperator*{\argmax}{argmax}
\DeclareMathOperator*{\argmin}{argmin}

\usepackage[colorinlistoftodos]{todonotes}

\tikzset{
  rednode/.style = {rectangle, draw=black!60, fill=red!20, very thick, minimum size=5mm,outer sep=0pt},
  bluenode/.style = {rectangle, draw=black!60, fill=blue!20, very thick, minimum size=5mm,outer sep=0pt},
  greennode/.style = {rectangle, draw=black!60, fill=green!20, very thick, minimum size=5mm,outer sep=0pt},
  whitenode/.style = {rectangle, draw=black!60, fill=white!20, very thick, minimum size=5mm,outer sep=0pt},
}

\newtheorem{example}{Example}

\title{Learning Optimal Temperature Region for Solving \\Mixed Integer Functional DCOPs\footnote{\copyright\ International Joint Conferences on Artificial Intelligence (IJCAI), all rights reserved. Proc. of the 29th International Joint Conference on Artificial Intelligence. Link to paper: https://www.ijcai.org/Proceedings/2020/0038.pdf.}}

\author{
Saaduddin Mahmud$^1$\and
Md. Mosaddek Khan$^1$\and
Moumita Choudhury$^{1}$\and
Long Tran-Thanh$^2$\And
Nicholas R. Jennings$^3$
\affiliations
$^1$Department of Computer Science and Engineering, University of Dhaka\\
$^2$School of Electronics and Computer Science, University of Southampton\\
$^3$Departments of Computing and Electrical and Electronic Engineering, Imperial College London
\emails
saadmahmud14@gmail.com,
mosaddek@du.ac.bd,
moumitach22@gmail.com,\\
ltt08r@ecs.soton.ac.uk,
n.jennings@imperial.ac.uk
}

\begin{document}

\maketitle

\begin{abstract}
Distributed Constraint Optimization Problems (DCOPs) are an important framework for modeling coordinated decision-making problems in multi-agent systems with a set of discrete variables. Later works have extended DCOPs to model problems with a set of continuous variables, named Functional DCOPs (F-DCOPs). In this paper, we combine both of these frameworks into the Mixed Integer Functional DCOP (MIF-DCOP) framework that can deal with problems regardless of their variables' type. We then propose a novel algorithm $-$ Distributed Parallel Simulated Annealing (DPSA), where agents cooperatively learn the optimal parameter configuration for the algorithm while also solving the given problem using the learned knowledge. Finally, we empirically evaluate our approach in DCOP, F-DCOP, and MIF-DCOP settings and show that DPSA produces solutions of significantly better quality than the state-of-the-art non-exact algorithms in their corresponding settings. 
\end{abstract}
\section{Introduction}
Distributed Constraint Optimization Problems (DCOPs) are a widely used framework for coordinating interactions in cooperative multi-agent systems. More specifically, agents in this framework coordinate value assignments to their variables in such a way that minimizes constraint violations by optimizing their aggregated costs \cite{yokoo1998distributed}. DCOPs have gained popularity due to their applications in various real-world multi-agent coordination problems, including distributed meeting scheduling \cite{Maheswaran2004TakingDT}, sensor networks \cite{farinelli2014agent} and smart grids \cite{fioretto2017distributed}.\par

Over the last two decades, several algorithms have been proposed to solve DCOPs, and they can be broadly classified into two classes: exact and non-exact. The former always provide an optimal solution of a given DCOP. However, solving DCOPs optimally is NP-hard \cite{modi2005adopt}, thus scalability becomes an issue as the problem size grows. In contrast, non-exact algorithms compromise some solution quality for scalability. Among the non-exact algorithms, DSA \cite{zhang2005distributed}, DSAN \cite{ArshadDistributedSA}, MGM \& MGM2 \cite{Maheswaran2004Distributed}, Max-Sum \cite{farinelli2008decentralised,gene}, Max-Sum\_ADVP \cite{zivan2012max}, DSA-SDP \cite{zivan2014explorative}, GDBA \cite{okamoto2016distributed}, PD-Gibbs \cite{dgibbs} and AED \cite{Mahmud2020} are commonplace.\par

In general, the classical DCOP framework assumes that all the variables that are used to model a problem are discrete. However, many real-world applications (e.g. target tracking sensor orientation \cite{fitzpatrick2003distributed}, sleep scheduling of wireless sensors \cite{hsin2004network}) are best modelled with continuous variables. In order to address this, \citeauthor{stranders2009decentralised} \citeyear{stranders2009decentralised} proposed a framework that facilitates the use of continuous variables, later referred to as a Functional DCOP (F-DCOP) \cite{Choudhury2019APS}. In contrast to classical DCOP, F-DCOP assumes all the variables are continuous and models all the constraints in the form of functions of those variables (instead of tabular form in classical DCOP). Among the F-DCOP solvers,  CMS~\cite{stranders2009decentralised}, HCMS~\cite{voice2010hybrid}, D-Bay \cite{fransman2020distributed}, EF-DPOP \& AF-DPOP \cite{hoang2019new} and PFD~\cite{Choudhury2019APS} are well known.\par

Against this background, it is obvious that the classical DCOP and F-DCOP can only deal with problems having discrete and continuous valued variables, respectively. In this paper, we first combine both of them into a Mixed Integer Functional DCOP (MIF-DCOP) framework, that can deal with a problem regardless of its variable types and representation of the constraints. We then develop a new algorithm that we call Distributed Parallel Simulated Annealing (DPSA) that can be directly applied to DCOPs\:and\:F-DCOPs, and even more importantly to their generalized version MIF-DCOPs.\par

The DPSA algorithm is based on Simulated Annealing (SA) meta-heuristic which is motivated by a physical analogy of controlled temperature cooling (i.e. annealing) of a material \cite{Kirkpatrick1983OptimizationBS}. One of the most important factors that influence the quality of the solution produced by SA is this temperature parameter, widely denoted as $T$. More precisely, SA starts with a high value of T and during the search process continuously cools it down to near zero. When T is high, SA only explores the search space without exploiting. This makes its behaviour similar to a random search procedure. On the other hand, when T is near zero, SA tends to only exploit and thus the exploration capability demises. In such a scenario, SA emulates the behaviour of a greedy algorithm. In fact, SA most effectively balances between exploration and exploitation in some optimal temperature region that lies in between these two extremes. Several existing works also discuss a constant optimal temperature where SA performs the best \cite{CONNOLLY199093,alrefaei1999simulated}. Unfortunately, the optimal temperature region varies from one type of problem to another and from one instance to another of the same type problem (with different constraint function, constraint density, number of agents, etc.).\par
In light of the above, we introduce a novel method where agents cooperatively try to learn this optimal temperature region using a Monte Carlo importance sampling method called Cross-Entropy sampling \cite{Kroese2011HandbookOM} (discussed in the background). Using the learned knowledge during this process, agents also cooperatively solve the given problem. This results in a significant improvement of solution quality compared to the state-of-the-art algorithms in both DCOP and F-DCOP settings (see Section~\ref{ExpSec} for details). Moreover, we apply and evaluate both DSAN (i.e. the only other SA based DCOP solver) and DPSA in MIF-DCOP settings and show that DPSA outperforms DSAN in this setting by a\:notable\:margin. 


\section{Background}
We first formally define DCOPs and F-DCOPs which will be the basis for the MIF-DCOP. We then conclude this section with a brief review of the literature necessary for this work. 
\subsection{DCOP Framework}
A DCOP is defined by a tuple $ \langle X,D,F,A,\delta \rangle $ \cite{modi2005adopt}. $A$ is a set of agents $\{a_1, a_2, ..., a_n\}$. $X$ is a set of discrete variables $\{x_1, x_2, ..., x_m\}$, which are being controlled by the set of agents $A$. $D$ is a set of discrete and finite variable domains $\{D_1, D_2, ..., D_m\}$, where each $D_i$ is a set containing values which may be assigned to its associated variable $x_i$. $F$ is a set of constraints $\{f_1,f_2,...,f_l\}$, where $f_i \in F$ is a function of a subset of variables $x^i \subseteq X$ defining the relationship among the variables in $x^i$. Thus, the function $f_i : \times_{x_j \in x^i} D_j \to \!R $ denotes the cost for each possible assignment of the variables in $x^i$. $\delta: X \rightarrow A$ is a variable-to-agent mapping function which assigns the control of each variable $x_i \in X$ to an agent of $A$ \cite{nodeto}. Each variable is controlled by a single agent. However, each agent can hold several variables.\par 
Within the framework, the objective of a DCOP algorithm is to produce $X^*$; a complete assignment that minimizes\footnote{For a maximization problem, replace $\argmin$ with $\argmax$.} the aggregated cost of the constraints as shown in Equation~\ref{eqobj}.
\vspace{-2mm}
\begin{equation}
    X^* = \argmin_X \sum_{i=1}^{l} f_i(x^i)
    \label{eqobj}
\end{equation}
For ease of understanding, we assume that each agent controls one variable. Thus, the terms `variable' and `agent' are used interchangeably throughout this paper.

\subsection{Functional DCOP Framework}
F-DCOPs can be defined by a tuple $ \langle X,D,F,A,\delta \rangle$, where $A$ and $\delta$ have the same definition as in the DCOP framework. However, the set of variables, $X$ and the set of domains, $D$, are defined as follows: $X$ is a set of continuous variables $\{x_1,x_2,...,x_m\}$ that are controlled by agents in $A$. $D$ is a set of continuous domains $\{D_1, D_2,...,D_m\}$, where each variable $x_i$ can take any value between a range, $D_i$ = [$LB_i$, $UB_i$]. A notable difference between F-DCOP and DCOP is found in the representation of the cost function $F$. In DCOPs, the cost functions are conventionally represented in tabular form, while in F-DCOPs, each constraint is represented in the form of a function \cite{hoang2019new}.
\begin{algorithm}[t]
\DontPrintSemicolon
Initialize parameter vector $\theta$, $\#S$, $G$,  $\alpha$\;
\While {condition not met}{
    $X \leftarrow$ take $\#S$ samples from distribution $\mathcal{G}$($\theta$)\;
    $S \leftarrow$ Evaluate points in $X$ on the objective\;
    $X \leftarrow$ sort($X$, $S$)\;       
    $\theta_{new} \leftarrow$ calculate updated $\theta$ using X(1:G)\;
    $\theta \leftarrow $ $(1-\alpha)*\theta+\alpha*\theta_{new}$
}
\caption{Cross-Entropy Sampling}
\label{algo:CEM}
\end{algorithm}
\subsection{Distributed Simulated Annealing}
Distributed Simulated Annealing (DSAN) \cite{ArshadDistributedSA} is the only existing Simulated Annealing (SA) based DCOP solver. DSAN is a local search algorithm that executes the following steps iteratively:
\begin{itemize}
    \item Each agent $a_i$ selects a random value $d_j$ from domain $D_i$.
    \item Agent $a_i$ then assigns the selected value to $x_i$ with the probability $min(1,exp(\frac{\Delta}{t_i}))$ where, $\Delta$ is the local improvement if $d_j$ is assigned and $t_i$ is temperature at iteration $i$. Note that authors of DSAN suggest that the value of $t_i = \frac{Max\_Iteration}{i^2}$ or $t_i = \frac{1}{i^2}$. However, setting the value of the temperature parameter with such a fixed schedule does not take into account their impact on the performance of the algorithm. 
    \item Finally, agents notify neighbouring agents if the value of a variable changes.
\end{itemize}
\subsection{Anytime Local Search}
Anytime Local Search (ALS) is a general framework that gives distributed iterative local search DCOP algorithms such as DSAN described above an anytime property. Specifically, ALS uses a BFS-tree to calculate the global cost (i.e. evaluate Equation \ref{eqobj}) of the system's state during each iteration and keeps track of the best state visited by the algorithm. Hence, using this framework, agents can carry out the best decision that they explored during the iterative search process instead of the one that occurs at the termination of the algorithm (see  \cite{zivan2014explorative} for more details).   
\subsection{Cross-Entropy Sampling}
Cross-Entropy (CE) is a Monte Carlo method for importance sampling. CE has successfully been applied to importance sampling, rare-event simulation and optimization (discrete, continuous, and noisy problems) \cite{Kroese2011HandbookOM}. Algorithm \ref{algo:CEM} sketches an example that iteratively searches for the optimal value of the $X$ within a search space. The algorithm starts with a probability distribution $\mathcal{G}(\theta)$ over the search space with parameter vector $\theta$ initialized to a certain value (that may be random). At each iteration, it takes \#$S$ (which is a parameter of the algorithm) samples from the probability distribution $\mathcal{G}$($\theta$) (line $3$). After that, each sample point is evaluated on a problem dependent objective function. The top $G$ among the \#$S$ sample points are used to calculate the new value of $\theta$ which is referred to as $\theta_{new}$ (lines $4-6$). Finally, $\theta_{new}$ is used to update $\theta$ (line $7$). At the end of the learning process, most of the probability density of $\mathcal{G}(\theta)$ will be allocated near the optimal value of $X$.      
\section{Mixed Integer Functional DCOP Framework}
We now formulate Mixed Integer Functional DCOP (MIF-DCOP) that combines both the classical DCOP and F-DCOP. This removes the requirement of all the variables being either continuous or discrete and constraint being represented in tabular or functional form. More formally, an MIF-DCOP is a tuple $\langle A, X, D, F, \delta \rangle$, where $A$, and $\delta$ are as defined in standard DCOPs. The key differences are as follows:
\begin{itemize}
    \item $X$ is a set of variables  $\{x_1, x_2, ..., x_m\}$, where each $x_i \in X$ is either a discrete or a continuous variable.
    \item $D$ is a set of domains $\{D_1, D_2, ..., D_m\}$. If a variable $x_i$ is discrete, its domain $D_i$ is the same as it is in the DCOP framework; otherwise, $D_i$ is the same as it is in the F-DCOP model.
    \item $F$ is a set of constraint functions $\{f_1,f_2,...,f_l\}$. Each constraint function $f_i$ can be modeled as follows:
    when the subset of the variables involved with $f_i$ contains only discrete variables, it can be modeled either in tabular form or functional form. Otherwise it is modeled only in functional form. For example: Consider a binary constraint function $f(x_1,x_2)$ between two variables $x_1$ and $x_2$. Let $D_1 = [-1000, 1000]$ and $D_2 = \{red, green\}$. Example of $f(x_1,x_2)$ under MIF-DCOP is: $f(x_1,x_2) = I(x_2 = red) *f_{red}(x_1)+ I(x_2 = green)*f_{green}(x_2)$ where $I(.)$ is an identity function. 
    
\end{itemize}
\begin{figure}[t]
  \hspace*{-1cm}   
  \includegraphics[scale = 0.60]{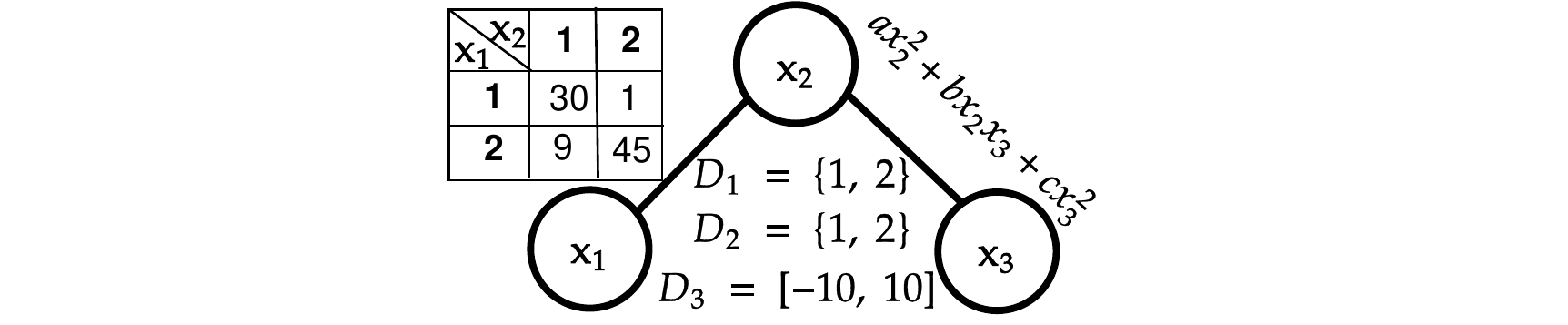}
  \caption{Example of a MIF-DCOPs.}
  \label{mif}
\end{figure}
It is worth highlighting that both DCOPs and F-DCOPs are special cases of MIF-DCOPs wherein either all the variables are discrete and constraints are in tabular form or all the variables are continuous and constraints are in functional form, respectively.\par 
MIF-DCOPs are specifically useful when the variables in $X$ represent decisions about a heterogeneous set of actions. Suppose, for instance, we want to model a target classification problem in a network of optical sensor agents. In the F-DCOP model of this problem \cite{Stranders2010DecentralisedCO}, an agent was only able to rotate their sensor to change its viewing direction. Now imagine that agents also can optically zoom in or zoom out to increase clarity or field of vision, respectively. The decision about rotation can be best modeled with a continuous variable (i.e. $rotation = [\ang{0}, \ang{360}]$, as described in \cite{Stranders2010DecentralisedCO}) and the decision about optical zoom is naturally modeled using discrete variables (e.g. $zoom = \{1x, 2x, 3x\}$). Other possible scenarios where MIF-DCOP can be applied are: mobile sensor applications where agents need to select both their location and the target they are covering and solving MIP problems in distributed settings. These problems, and many other similar problems where heterogeneous types of decision variables are needed, can easily be modelled with the newly proposed MIF-DCOPs. We provide a small example of MIF-DCOP in Figure~\ref{mif}.       

\section{The DPSA Algorithm}
\begin{algorithm}[!h]
\DontPrintSemicolon
\SetKwFunction{Update}{Update\_Parameter}
\SetKwFunction{Round}{Simulate}
\SetKwFunction{Temp}{Scheduler}
\SetKwFunction{Neighbour}{Select\_Next}
\SetKwFunction{Main}{Main}  
\SetKwProg{In}{Function}{:}{} 
 
   \In{\Round{isLearning}}{
      \eIf{isLearning is True}{
            set $x_{i,k}$ $\forall k \in \{1,...,K\}$ to same random value\;
            $L \leftarrow S_{len}$\;
        }
        {
            set $x_{i,k}$ $\forall k \in \{1,...,K\}$ to value of $x_i$\;
            $L \leftarrow Itr_{max} - R_{max}*S_{max}*S_{len}$\;
        }
        \For{$l = 1 ... L$}{
            send value of $x_{i,k}$ $\forall k \in \{1,...,K\}$ to neighbouring agents\;
            receive value of $x_{i,k}$ $\forall k \in \{1,...,K\}$ from neighbouring agents\;
            Update $x_i$ using Modified\_ALS($\cdot$)\;
             \For{$k = 1 ... K$}{
       	        $v \leftarrow $\Neighbour{$D_i$}\;
       	        $t_k \leftarrow $\Temp{$l$,$k$,isLearning}\;
       	        Set $x_{i,k}$ to $v$ with probability $min(1,\exp(\frac{\Delta_k}{t_k}))$\;
           }
        }    
   }
   
   \In{\Update{$\theta,E,T$}}{
        $Threshold \leftarrow $G-th best of set $E$\; 
        $SelectedSample \leftarrow \{t_k: e_k\le Threshold\}$\;
        Update $\theta$ using $SelectedSample$\;
   }
   \In{\Main{}}{
       Construct BFS Tree\;
       Initialize parameters: Parameter Vector $\theta$, $Itr_{max}, R_{max}, S_{max}, S_{len}, G, K$\;
        \For{$R = 1 ... R_{max}$ AND Conditions are met}{
            $E \leftarrow \{e_1 = 0,e_2 = 0,...,e_K = 0\}$\;
            \eIf{the agent is the root}
           	    {
           	        $T \leftarrow \{t_1,t_2,...,t_K\}$sampled from $\mathcal{G}(\theta)$\;
           	    }
           	    {
           	        $T \leftarrow$ Receive T from the parent agent in the BFS-Tree\;
           	    }
           	    Send $T$ to all the child agents in BFS-Tree\;
             \For{$s = 1 ... S_{max}$}{
           	    Synchronously start \Round(True)\;
           	    Wait for Modified\_ALS($\cdot$) to terminate\;
           	    \For{$e_k \in E$}
           	    {
           	        $e_k \leftarrow e_k + \frac{bestCost_{S,k}}{S_{max}}$
           	    }
               }
             \Update($\theta,E,T$)\;
        }
        Synchronously start \Round(False)\;
   }

\caption{The DPSA Algorithm}
\label{algo:DPSA}

\end{algorithm}
We will now describe the DPSA algorithm for solving MIF-DCOPs (Algorithm $2$\footnote{Select\_Next(.), Scheduler(.) and  Modified\_ALS($\cdot$) are described in text.}). As discussed earlier, a big motivation behind DPSA is to learn the optimal temperature region for Simulated Annealing (SA). Thus, it is important that we formally define optimal temperature (Definition 1) and optimal temperature region (Definition 2).\par
\textbf{Definition 1.} \textit{An Optimal Temperature given simulation length $L$} is a constant temperature at which the expected solution cost yielded by running SA for L iterations is the lowest of all the temperatures $> 0$.\par   
\textbf{Definition 2:} \textit{An Optimal Temperature Region (OTR) of length $\epsilon$} is a continuous interval $[T_{min},T_{max}]$ where $T_{max} - T_{min} \le \epsilon$ and contains the optimal temperature. If we set $T_{min}$ to near zero and $T_{max}$ to a very large number, it will always be an OTR by the above definition; although not a useful one. The proposed DPSA algorithm tries to find an OTR with sufficiently small $\epsilon$.\par 
The DPSA algorithm consists of two main components: the parallel SA component and the learning component. The parallel SA component (lines $1-15$), is an extension of the existing DSAN algorithm that simulates $K$ systems in parallel. Additionally, we introduce the Select\_Next($\cdot$) function in which an agent uses different strategies to select a value from its domain depending on its variable type, Continuous or Discrete (line $13$). This simple modification makes DPSA (also DSAN) applicable to DCOP, F-DCOP and MIF-DCOP. We also modify the existing ALS framework to make it compatible with parallel simulation. \par
The other significant component of DPSA is the iterative learning component sketched in the pseudo-code from line $16$ to $36$. It starts with a large temperature region and iteratively tries to shrink it down to an OTR of a small length ($\epsilon$). To obtain this, at each iteration, agents cooperatively perform actions (i.e. synchronously simulate parallel SA with different constant temperatures), collect feedback (i.e. the cost yields by different simulations) and use the feedback to update the current temperature region toward the goal region. The underlining algorithm that is used in the learning process is based on cross-entropy importance sampling. However, to make DPSA sample efficient, we present modifications that significantly reduce the number of iterations and parallel simulations needed.\par 
We structure the rest of our discussion in this section as follows: we first describe parallel SA and our modification of ALS. Then we discuss the learning component of DPSA and the techniques to make it sample efficient for DPSA. Finally, we analyze the complexity of DPSA.\par  

In DPSA, the Simulate($\cdot$) function (lines $1-15$) runs SA on $K$ copies of a system in parallel. This function is called in two different contexts: during learning to collect feedback (line $31$) and after the learning process has ended (line $36$). The main difference is that in the first context the function runs a short simulation in $K$ different constant temperatures (one fixed temperature for each copy). In the second case, the function runs for significantly longer and all $K$ systems run on the learned OTR with a fixed scheduler (discussed shortly). In addition, in the first case, all copies are initialized with the same random value from the domain. This is done because we want to identify the effect of different constant temperatures on the simulation step, and initializing them with different initial states would add more noise to the feedback. In the second case, we initialize with the best state found so far. Note that, to avoid confusion, we use $x_i$ to refer to the actual decision variable and $x_{i,k}$ to refer to $x_i$ on the k-th copy of the system. The parameter $isLearning$ (line $2$) is used to represent the context in which the function was called. Depending on its value, variables of all K systems are initialized and the length of the simulation is set (lines $2-7$) ($S_{len}$ is the simulation length during learning). After that, the main simulation starts and runs for $L$ iterations.\par 
\begin{figure}[t]
  \hspace*{-1cm} 
  \includegraphics[scale = 0.60]{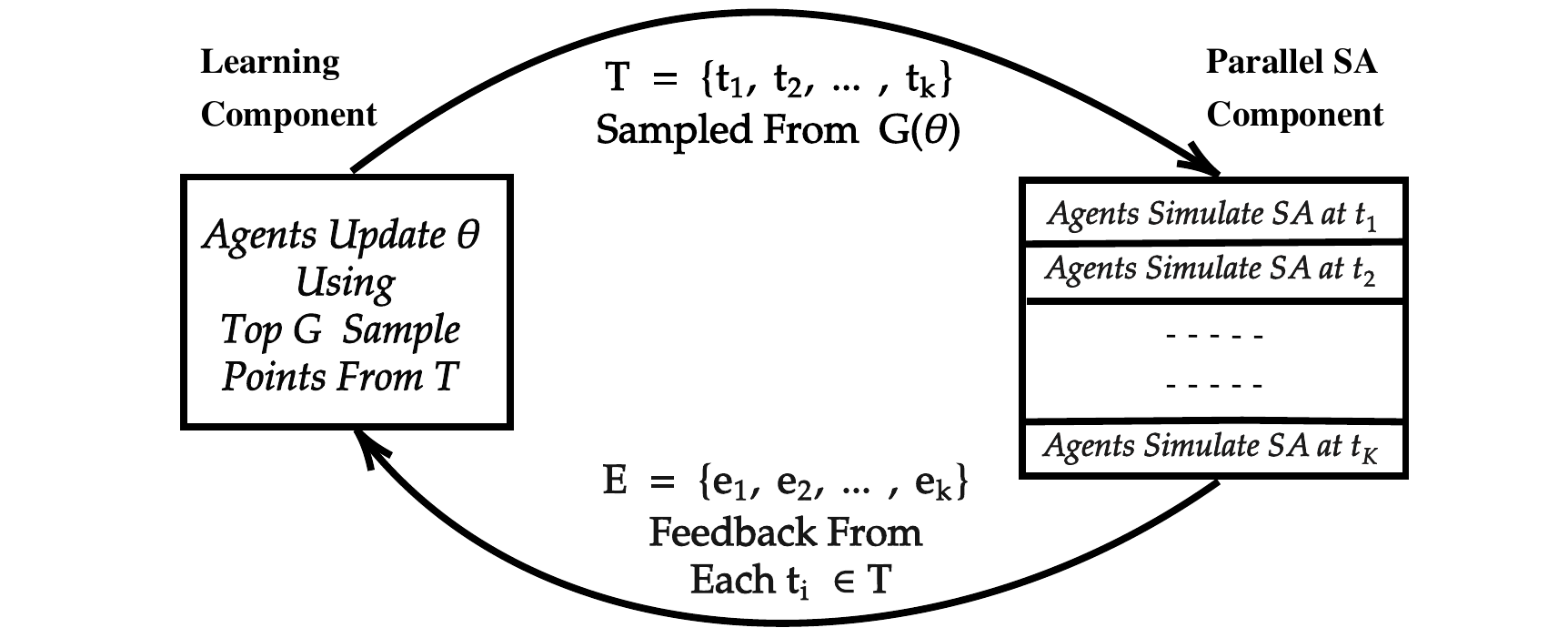}
  \caption{Overview of DPSA Algorithm}
  \label{ov}
\end{figure}
At the start of each iteration, each agent $a_i$ shares the current state of each system (i.e. the variable assignment of each system) with their neighbours (lines $9-10$). Each agent $a_i$ then updates $x_i$ and performs other operations related to Modified\_ALS($\cdot$) (discussed shortly) (line $11$). After that, for each of the K systems, agent $a_i$ picks a value from its domain $D_i$ (line $13$) by calling the function $Select\_Next(\cdot)$. If $D_i$ is discrete, it picks this value uniform randomly. If $D_i$ is continuous, it picks it either using a Gaussian distribution $\mathcal{N}(x_{i,k},\sigma)$ or a Uniform distribution $\mathcal{U}(LB_i,UB_i)$. Here, $UB_i$ and $LB_i$ is the bound of $D_i$. The difference is that Gaussian gives preference to a nearby value of the current assigned value, while uniform does not have any preference. Afterward, each agent selects the temperature for the current iteration i.e. $t_k$ (line $14$) by calling the function Scheduler($\cdot$). If it is called during the learning context, it is always set to a constant. More precisely, it is set to the k-th value of $T$ (from lines 26, 29). Otherwise, if the learned OTR is $[T_{min},T_{max}]$; it can be used with a temperature scheduler for example linear scheduler. To calculate the temperature using linear scheduler we use Equation \ref{lin}.
\begin{equation}
    \label{lin}
    T_{min}+(T_{max} - T_{min})\frac{L-l}{L}
\end{equation}

Finally, agents assign the value $v$ to $x_{i,k}$ with the probability $min(1,\exp(\frac{\Delta_k}{t_k}))$ where $\Delta_{k}$ is the local gain (i.e. improvement of the aggregated cost of the local constraints if the assignment is changed) of the k-th system (line $15$). If this gain is non-negative, it will always be changed (since $\exp(\frac{\Delta_k}{t_k}) \ge 1$). Otherwise, it will be accepted with a certain probability less than one.\par  
We now describe our modification of ALS. This is used to collect feedback from the simulations during learning and to give DPSA its anytime property. We modify ALS in the following ways:
\begin{itemize}
    \item Since DPSA simulated K systems in parallel, Modified\_ALS($\cdot$) keeps track of the best state and the cost found by each system separately within the duration of a call of Simulate($\cdot$) function. This is used for the feedback. 
    \item Modified\_ALS($\cdot$) also keeps track of the best state and cost across all K systems and all the calls of the Simulate($\cdot$) function. Using this, agents assign values to their decision variables. This is used to give DPSA its anytime property. 
\end{itemize}

The first part of the modification can easily be done by running each system at each call with its separate ALS. For the second part, we can have a meta-ALS that utilizes information of the ALSs in the first part to calculate the best state and cost across calls and systems.\par 
We now discuss the learning component (lines $16-36$). Here, we start with a probability distribution $\mathcal{G}(\theta)$ over a large temperature region $[T_{min}, T_{max}]$ where $\theta$ is the parameter vector of the distribution. At the start of each iteration, the root agent samples K points (i.e. a set of constant temperatures $T$ from this distribution (line $26$)). Agents then propagate this information down the pseudo-tree (lines $28-29$). After that, agents synchronously call the Simulate($\cdot$) function for $S_{max}$ times (line $31$). At each call, agents simulate SA in the K sampled constant temperatures (i.e. set T) in parallel. Then using Modified\_ALS($\cdot$), agents collect feedbacks i.e. cost of the best solution found by the simulation (line $32$). 
Then agents take a mean over $S_{max}$ feedback (lines $33-34$). This average should be an unbiased estimation of the expected solution quality i.e. the actual feedback given large $S_{max}$. Note that in the pseudo-code, we use $bestCost_{S,k}$ to refer to the best cost found by the k-th system in the S-th call. After all the feedback is collected, we use it to update the parameter vector using the Update\_Parameter($\cdot$) function (line $35$). The Update\_Parameter($\cdot$) function takes the G best sample point (lines $17-18$) and uses them to update the parameter vector (line $19$). In this way, agents iteratively learn the parameter vector $\theta$.\par 
The parameter vector $\theta$ and its update depends on the particular distribution $\mathcal{G}(\cdot)$ used. In this paper, we focus on two particular distributions namely Gaussian $\mathcal{N}(\cdot)$  and uniform $\mathcal{U}(\cdot)$. The parameter vector for $\mathcal{N}(\cdot)$ is  $\theta = [\mu, \sigma]$ and consists of the mean and the standard deviation. The new parameter vector is calculated as:
\begin{equation}
    \theta_{new} = [\mu(SelectedSample), \sigma(SelectedSample)]
\end{equation}{}
The parameter vector for $\mathcal{U}(\cdot)$ is  $\theta = [T_{min}, T_{max}]$ and consists of the current bound of temperature region. The new parameter vector is calculated as:
\begin{equation}
\small
    \theta_{new} = [min(SelectedSample), max(SelectedSample)]
\end{equation}{}
Finally, $\theta$ is updated as shown in Equation \ref{upd} where $\alpha$ is the learning rate. 
\begin{equation}
\label{upd}
    \theta = (1 - \alpha) * \theta + \alpha * \theta_{new}
\end{equation}{}
Updating parameters in the way discussed above requires a considerable amount of iterations and samples. To reduce the number of iterations and samples required, we now discuss a few techniques that we used in the experimental section. First, when the number of parallel simulations is small (i.e. the value of K is small) taking a random sample in a large range is not efficient, and taking a stratified sample will cover a large range better. For example, when using $\mathcal{U}(\cdot)$, we may take samples at regular interval using Equation \ref{uni}:
\begin{equation}
    \label{uni}
    T_{min}+(T_{max}-T_{min})*\frac{k-1}{K-1};  k \in \{1,2,...,K\}
\end{equation} 
Second, when $S_{max}$ is small, the estimation of expected cost becomes noisy. To address this, when two sample points produce feedback within a bound of each other, we consider them equally good. We calculate this bound $\gamma$ using Equation~\ref{sen}. 
\begin{equation}
\label{sen}
    \gamma = \mathcal{S} * bestCost_{*}
\end{equation}{}
Here, $\mathcal{S}$ stands for sensitivity and is an algorithm parameter and we use $bestCost_*$ to refer to the actual best cost found so far across all the calls of the Simulate(.) function. According to this, we may calculate the $Threshold$ in line 19 as follows: $Threshold \leftarrow$ G-th best of $E + \gamma$.\par    
Finally, when $R_{max}$ is small, setting learning rate to a larger value will speed up the learning process. However, if it is set too high, the algorithm might prematurely converge or skip the optimal temperature. Additionally, we can terminate before $R_{max}$, if all the sample points are within $\gamma$ of each other. We now provide an example of the learning process:
\begin{example}
Suppose, we have $\alpha = 0.4, G = 3, K = 10, \theta = [0.1,100]$ and we use a uniform distribution. In the first round, the sampled points will be (when taken using the regular interval):
$$
    T = [0.1,11.1,22.2,33.3,44.4,55.5,66.6,77.7,88.8,100]
$$
Let the feedback from each point be:
$$
    E = [50,40,30,25,32,42,57,70,95,130]
$$
The selected sample points (top $G = 3$ points) will be:
$$
    SelectedSample = [22.2,33.3,44.4]
$$
Finally, the parameter update will be (min and max of $SelectedSample$):
$$
    \theta_{new} = [22.2,44.4]
$$
$$
    \theta = 0.6*[0.1,100]+0.4*[22.2,44.4] = [8.9,77.8]
$$
This process will repeat until the termination conditions are met. We give a visual overview of the process in Figure \ref{ov}.
\end{example}
After the learning process ends, agents call the Simulate($\cdot$) function for the final time at line $36$. At this time, the simulation usually runs for longer on the learned optimal temperature region. This concludes our discussion on the learning component. It is worth noting that although we apply this learning component to learn parameter value for SA, it can also be applied to learn parameter(s) for other DCOP algorithms. 
In this way, it can be thought of as a generic distributed parameter learning algorithm for DCOPs.\par
In terms of complexity, the main cost is yielded by the Simulate($\cdot$) function. Each iteration of this function requires calculating the local gain $\Delta_k$ for $K$ systems. The calculation of local gain requires $O(|N|)$ complexity where $|N|$ is the number of the neighbours. Hence, the computation complexity is $O(K|N|)$ (per iteration and per agent). In terms of communication complexity, $K$ variable assignments are transferred at each iteration which gives it $O(K)$ complexity. Finally, agents have to save $K$ local variable assignments, each of which requires $O(|N|)$ memory, meaning the total memory requirement will be $O(K|N|)$. It is worth mentioning that the memory requirement of Modified\_ALS($\cdot$) is $O(K|H|)$ where $|H|$ is the height of the BFS\:tree. In Modified\_ALS($\cdot$), while the number of messages remains the same as ALS size of each message increase by an factor of $K$.   

\section{Experimental Results}\label{ExpSec}

We start by evaluating DPSA against the state-of-the-art DCOP algorithms on 7 different benchmarks. We then test DPSA and DSAN against the state-of-the-art F-DCOP solvers. Finally, we present comparative solution quality produced by DPSA and DSAN on a MIF-DCOP setting.\par
For the former, we consider the following seven benchmarking DCOP algorithms: DSA-C (p = 0.8), MGM-2 (offer probability 0.5), Max-Sum ADVP, DSA-SDP (pA = 0.6, pB = 0.15, pC = 0.4, pD = 0.8), DSAN, GDBA and PD-Gibbs. For all the benchmarking algorithms, the parameter settings that yielded the best results are selected. We compare these algorithms on six Random DCOP settings and Weighted Graph Coloring Problems (WGCPs). For all settings, we use Erd{\H{o}}s-R{\'e}nyi topology (i.e. random graph) to generate the constraint graphs \cite{erdHos1960evolution}. For random DCOPs, we vary the density $p$ from 0.1 to 0.6 and the number of agents from 25 to 75. For WGCPs, we set $p = 0.05$ and number of agents to 120. We then take constraint costs uniformly from the range $[1, 100]$ and set domain size to 10. For all the benchmarks, we use the following parameters for DPSA $Itr_{max} = 2500, R_{max} = 12, S_{max}=1, S_{lan} =100, \alpha =0.5, \mathcal{S} = .01$ and $K=16$. When selecting values for these parameters note that increasing $Itr_{max}, R_{max}, S_{max}, S_{lan}$ and $K$ will in general increase accuracy of the learning component in exchange for computational effort. A similar case is the learning rate $\alpha$ where decreasing it will increase accuracy but will slow down the learning process. Finally, for large values of $S_{max}$, sensitivity $\mathcal{S}$ will become less important. In all the settings of this section, DPSA uses $\mathcal{U}(\cdot)$ for CE. DPSA initializes parameter vector $\theta$ with a large temperature region $[10^{-3}, 10^3]$ for discrete settings and $[10^{-4}, 10^4]$ for continuous and mixed settings. In all of the settings described above, we run the benchmarking algorithms on 50 independently generated problems and 50 times on each problem for 500 ms. In order to conduct these experiments, we use a GCP-n2-highcpu-64 instance, a cloud computing service which is publicly accessible at cloud.google.com. Note that unless stated otherwise, all differences shown in this section are statistically significant for $p-value<0.01$. \par

\renewcommand{\arraystretch}{1}
\begin{table}[t]
\centering
\setlength{\tabcolsep}{2.5pt}
\begin{tabular}{|c|c |c|c| c|c |c|}
\hline
\textbf{\tiny Algorithm} & \multicolumn{2}{c|}{\textbf{\tiny $|A|$ = 25}} & \multicolumn{2}{c|}{\textbf{\tiny $|A|$ = 50}} & \multicolumn{2}{c|}{\textbf{\tiny $|A|$ = 75}} \\ \cline{2-7} 
                           & \textbf{\tiny P = 0.1}         & \textbf{\tiny P = 0.6}         & \textbf{\tiny P = 0.1}         & \textbf{\tiny P = 0.6}         & \textbf{\tiny P = 0.1}         & \textbf{\tiny P = 0.6}         \\ \hline
\tiny DSA-C                      & \tiny 432             & \tiny 5725            & \tiny 2605            & \tiny 27163           & \tiny 7089            & \tiny 65519           \\ \hline
\tiny DSA-SDP                    & \tiny 325             & \tiny 5635            & \tiny 2365            & \tiny 27210           & \tiny 6711            & \tiny 65600           \\ \hline
\tiny GDBA                       & \tiny 386             & \tiny 5465            & \tiny 2475            & \tiny 26950           & \tiny 6867            & \tiny 65156           \\ \hline
\tiny MGM-2                      & \tiny 352             & \tiny 5756            & \tiny 2481            & \tiny 27421           & \tiny 6962            & \tiny 65988           \\ \hline
\tiny PD-Gibbs                   & \tiny 398             & \tiny 5875            & \tiny 2610            & \tiny 27350           & \tiny 7178            & \tiny 65650           \\ \hline
\tiny MS\_ADVP                   & \tiny 400             & \tiny 5805            & \tiny 2550            & \tiny 27400           & \tiny 7058            & \tiny 66008           \\ \hline
\tiny DSAN                       & \tiny 408             & \tiny 5802            & \tiny 2639            & \tiny 27413           & \tiny 7224            & \tiny 66085           \\ \hline
\textbf{\tiny DPSA}             & \textbf{\tiny 268}    & \textbf{\tiny 5358}   & \textbf{\tiny 2136}   & \textbf{\tiny 26240}  & \textbf{\tiny 6276}   & \textbf{\tiny 63998}           \\ \hline
\end{tabular}

\caption{Comparison of DPSA and the benchmarking algorithms on difference configuration of random DCOPs.}
\label{TRG}
\end{table}

\begin{figure}[t]
\centering
  \includegraphics[scale = 0.55]{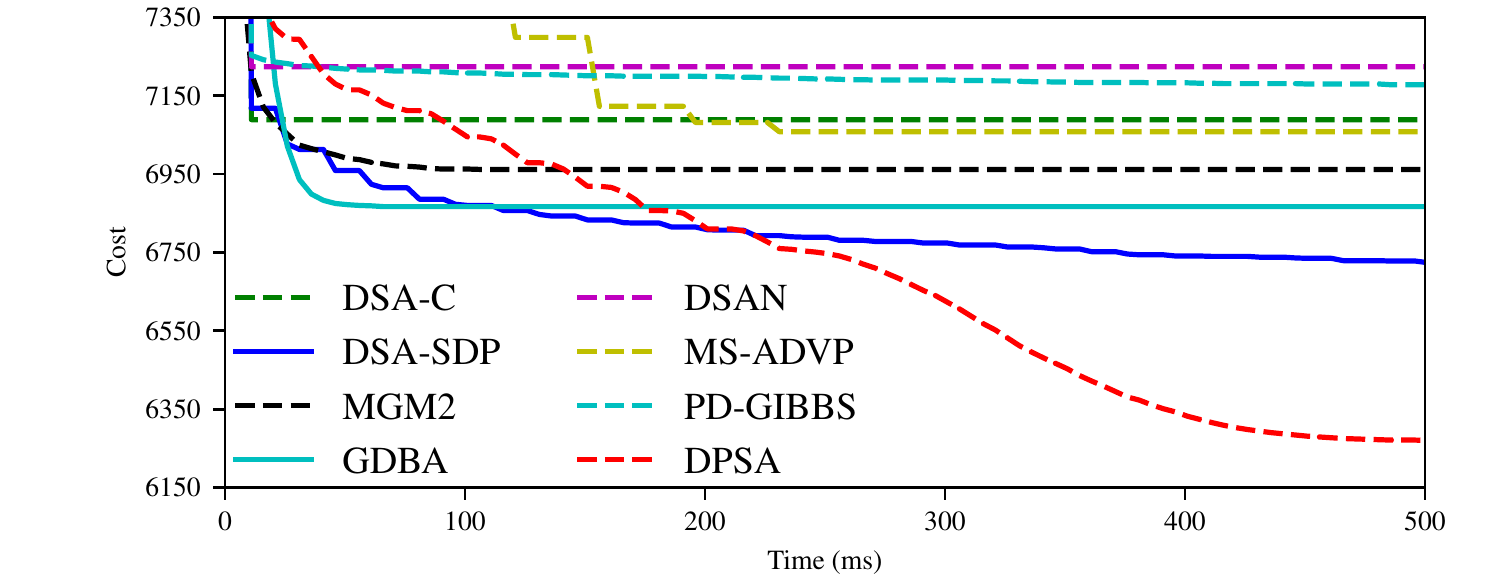}
  \vspace{-2mm}
  \caption{Comparison of DPSA and the benchmarking algorithms on random DCOPs ($|A|$ = 75, P = 0.1).}
  \label{R10}
\end{figure}

Figure~\ref{R10} shows a comparison between DPSA and the benchmarking algorithms on the random DCOPs ($|A|$ = 75 and p=0.1) setting. While Table~\ref{TRG} presents how performances of these algorithms vary with the number of agents and density. When the density is low, the closest competitor to DPSA is DSA-SDP. Even though both of the algorithms keep on improving the solution until the end of the run, DPSA makes significant improvement when it starts running in the optimal temperature region after the learning process ends, and we see a big decline after 250 ms. From the results in Table~\ref{TRG}, it can be observed that DPSA produces solutions that are $21\% - 6.7\%$ better than DSA-SDP depending on the number of agents. However, when the density is high ($p=0.6$), GDBA is the closest competitor to DPSA. In dense settings, DPSA outperforms GDBA by $1.8\% - 1.9\%$. Other competing algorithms perform equal or worse than GDBA and DSA-SDP and produce even bigger performance difference with DPSA (up to 15\% - 61\% in sparse settings and 9.6\% - 3.2\% in dense settings). Also note that the optimal cost for ($|A|$ = 25, p = 0.1), which we generate using the well-known DPOP \cite{Petcu2005ASM} algorithm, is 253, while DPSA produces 268 in the same\:setting.\par    
\begin{figure}[t]
\centering
  \includegraphics[scale = 0.55]{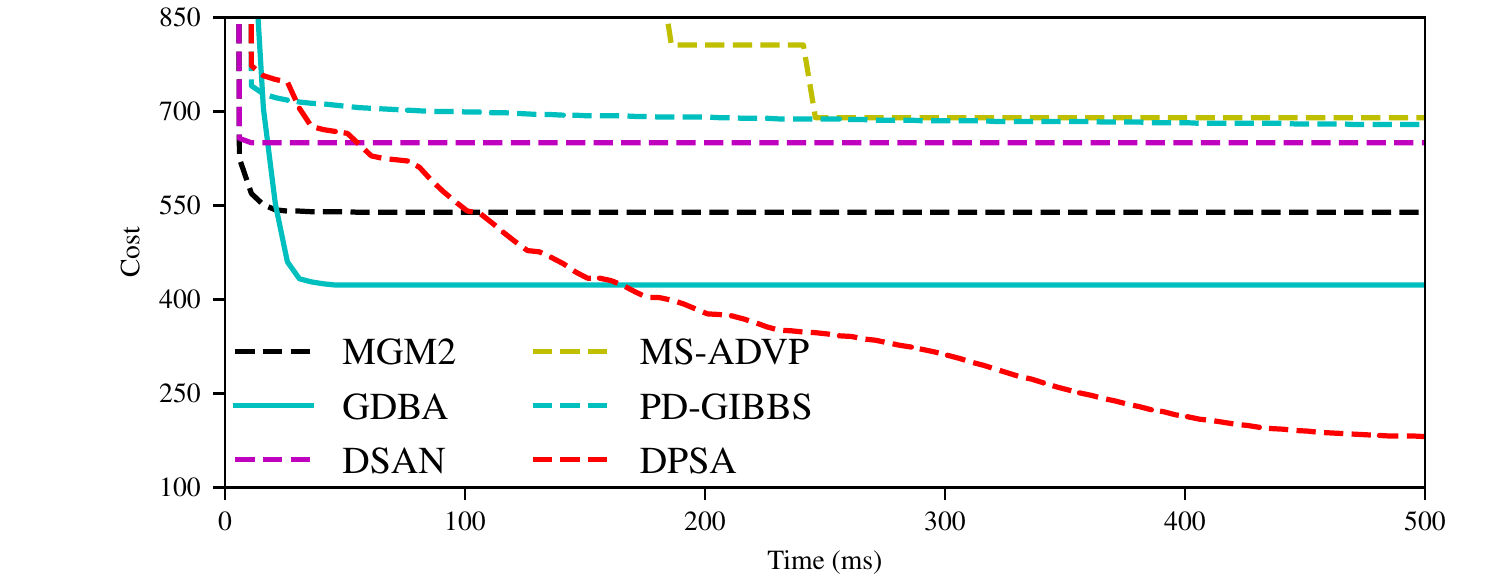}
  \vspace{-2mm}
  \caption{Comparison of DPSA and the benchmarking algorithms on weighted graph coloring problems ($|A|$ = 120, P = 0.05).}
  \label{W05}
\end{figure}
Figure \ref{W05} shows a comparison between DPSA and the benchmarking algorithms on the WGCPs ($|A|$ = 120 and p=0.05) benchmark. We see a similar trend here as observed in the random DCOP settings. For the first 1200 iterations (up to 250 ms) i.e. during the learning stage, DPSA improves the solution with several small steps, and after that, it takes a big step toward a better solution when ran longer in the OTR. In this experiment, DPSA demonstrates its notable performance. Among the benchmarking algorithms, GDBA is the closest but is still outperformed by DPSA by $1.33$ times. Among the other algorithms, DPSA finds solutions that are $3.65-1.95$ times better ($3.65$ times better than DSA-C). From the trend seen in Figures~\ref{R10} and \ref{W05} and performance produced by DPSA compared to the current state-of-the-art DCOP algorithms signifies that DPSA applied in the optimal temperature region is extremely effective at solving DCOPs. Since both DPSA and DSAN apply the same principle; the big performance gain of DPSA in terms of solution quality can be credited to the fact that DPSA runs significantly longer near the optimal temperature.\par
\begin{figure}[t]
\centering
  \includegraphics[scale = 0.55]{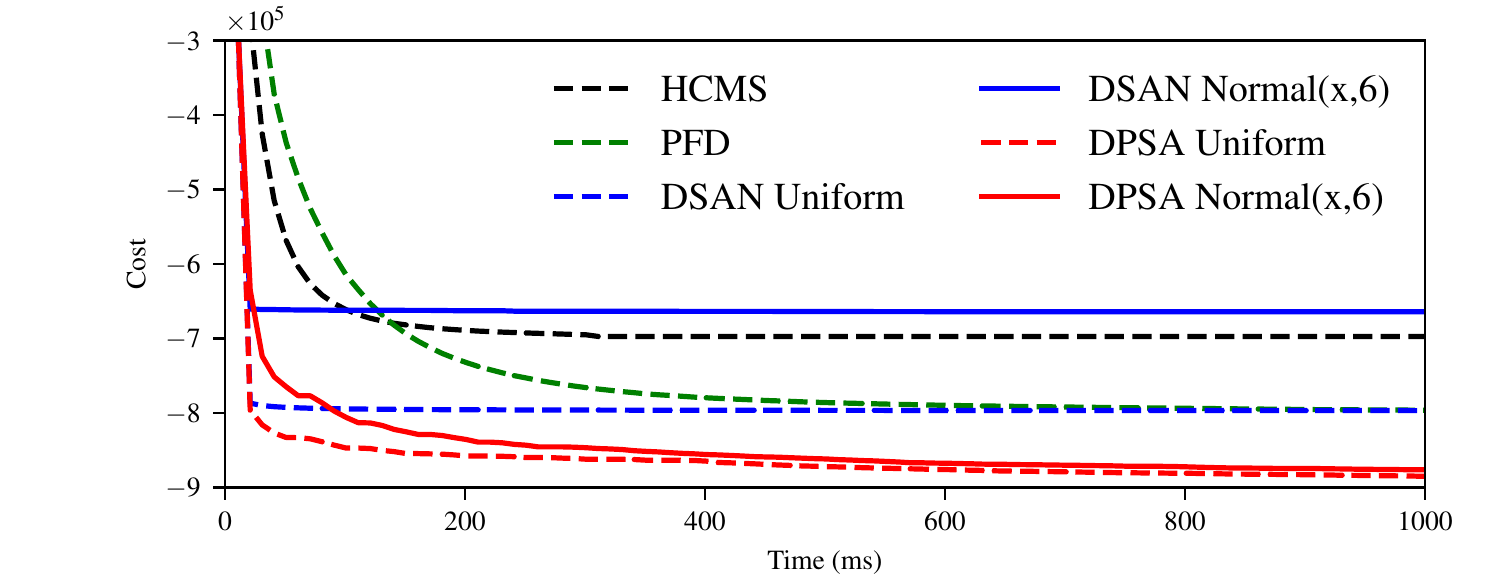}
  \vspace{-2mm}
  \caption{Comparison of DPSA and the benchmarking algorithms on binary quadratic F-DCOPs ($|A|$ = 50, P = 0.2).}
  \label{F50}
\end{figure}
We now compare DPSA and DSAN against current state-of-the-art F-DCOP solvers namely: PFD and HCMS on a large random graph with binary quadratic functions of the form $ax^2 + bxy + cy^2$. To generate random graphs, we use  Erd{\H{o}}s-R{\'e}nyi topology with number of agents set to 50 and $p = 0.2$. We choose coefficients of the cost functions (a, b, c) randomly between $[-5, 5]$ and set the domains of each agent to $[-50, 50]$. We run all algorithms on 50 independently generated problems and 50 times on each problem for 1 second and use the same hardware setup as the previous settings. For PFD, we use the same configuration suggested in \cite{Choudhury2019APS}. For HCMS, we choose the number of discrete points to be 3. The discrete points are chosen randomly between the domain range. Finally, we use following parameters for DPSA $Itr_{max} = 3000, R_{max} = 12, S_{max}=1, S_{lan} =120, \alpha =0.5, \mathcal{S}=0.005$ and $K=25$. To select neighbours in DPSA and DSAN, we use both uniform distribution and Normal distribution with $\sigma = 6$ over the domain.\par

Figure \ref{F50} shows a comparison between DPSA and the benchmarking algorithms on the binary quadratic F-DCOP ($|A|$ = 50, P = 0.2) benchmark. For this benchmark, uniform distribution for neighbour selection performs better than normal distribution both for DPSA and DSAN. DSAN (Uniform) produces similar solution quality as PFD. However, it has significantly improved anytime performance. On the other hand, DPSA produces solutions of significantly improved quality with the closest competitor PFD and DSAN being outperformed by 10.1\%. This demonstrates that DPSA is also an effective F-DCOPs solver.\par 
\begin{figure}[t]
\centering
  \includegraphics[scale = 0.55]{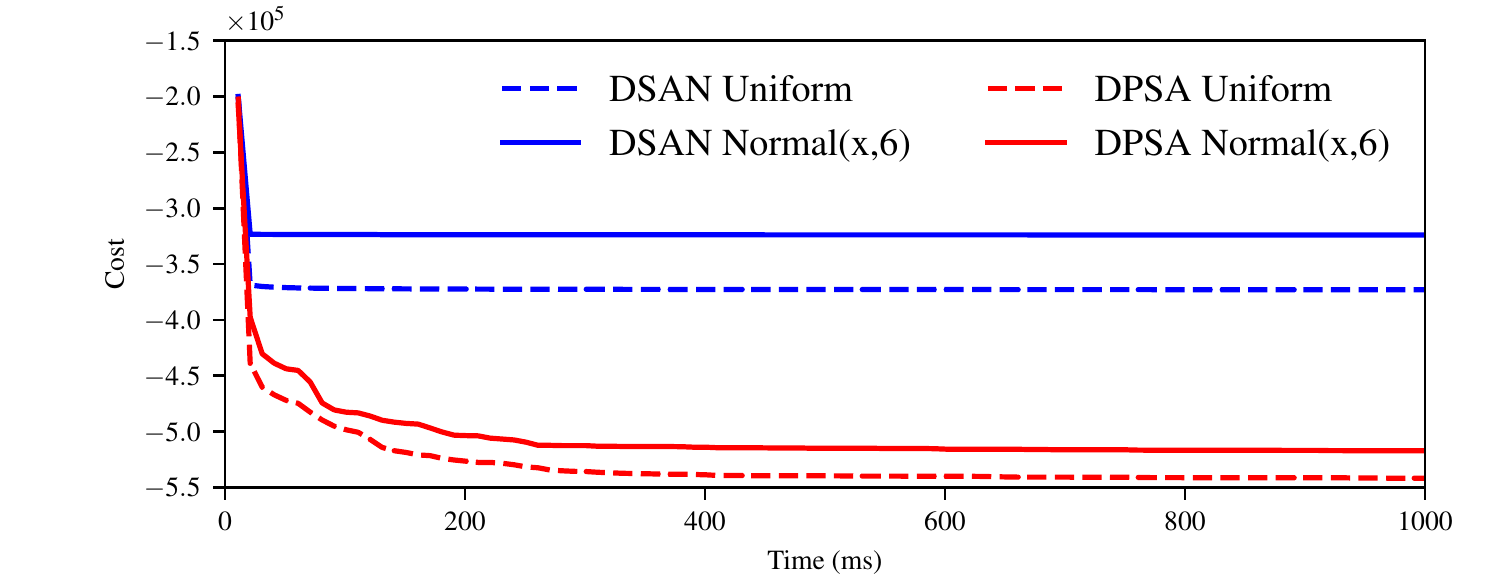}
  \vspace{-3mm}
  \caption{Comparison of DPSA and the benchmarking algorithms on binary quadratic MIF-DCOPs ($|A|$ = 50, P = 0.2).}
  \label{MIF50}
\end{figure}
Finally, we test DPSA and DSAN in the MIF-DCOP setting. For this, we use the same set of problems as F-DCOPs except that we randomly make 50\% of the variables discrete and set their domain to $\{-20, ... , 20\}$. Figure~\ref{MIF50} shows anytime performance of DPSA and DSAN on the binary quadratic MIF-DCOP ($|A|$ = 50, P = 0.2) benchmark. We see a similar trend as we have seen in F-DCOPs benchmark. Here, DSAN converges fast to local optima and fails to make any further improvement. On the other hand, DPSA avoids local optima through maintaining a good balance between exploration and exploitation by operating in the optimal temperature region and produces a 46\% better solution.\par

\section{Conclusions}
In this paper, we introduce MIF-DCOP framework that generalizes the well-known DCOP and F-DCOP models. We then propose a versatile algorithm called DPSA that can be applied to DCOPs, F-DCOPs and MIF-DCOPs. Finally, our empirical results depict that DPSA outperforms the state-of-the-art DCOP and F-DCOP algorithms by a significant margin and produces high quality solution compared to DSAN in MIF-DCOPs. In future, we intend to apply the learning component of DPSA to other DCOP algorithms such as DSA-SDP and Max-sum with damping to improve their solution quality. 

\bibliographystyle{named}
\balance
\bibliography{ijcai20}

\begin{thebibliography}{}

\bibitem[\protect\citeauthoryear{Alrefaei and
  Andrad{\'o}ttir}{1999}]{alrefaei1999simulated}
M.~H. Alrefaei and S.~Andrad{\'o}ttir.
\newblock A simulated annealing algorithm with constant temperature for
  discrete stochastic optimization.
\newblock {\em Management science}, 45:748--764, 1999.

\bibitem[\protect\citeauthoryear{Arshad and
  Silaghi}{2004}]{ArshadDistributedSA}
M~Arshad and M.~C. Silaghi.
\newblock Distributed simulated annealing.
\newblock 2004.

\bibitem[\protect\citeauthoryear{Choudhury \bgroup \em et al.\egroup
  }{2020}]{Choudhury2019APS}
M.~Choudhury, M.~Mahmud, and M.M. Khan.
\newblock A particle swarm based algorithm for functional distributed
  constraint optimization problems.
\newblock In {\em AAAI}, 2020.

\bibitem[\protect\citeauthoryear{Connolly}{1990}]{CONNOLLY199093}
D.T. Connolly.
\newblock An improved annealing scheme for the qap.
\newblock {\em European Journal of Operational Research}, 46:93 -- 100, 1990.

\bibitem[\protect\citeauthoryear{Erd{\H{o}}s and
  R{\'e}nyi}{1960}]{erdHos1960evolution}
P.~Erd{\H{o}}s and A.~R{\'e}nyi.
\newblock On the evolution of random graphs.
\newblock {\em Publ. Math. Inst. Hung. Acad. Sci}, 5(1):17--60, 1960.

\bibitem[\protect\citeauthoryear{Farinelli \bgroup \em et al.\egroup
  }{2008}]{farinelli2008decentralised}
A.~Farinelli, A.~Rogers, A.~Petcu, and N.R. Jennings.
\newblock Decentralised coordination of low-power embedded devices using the
  max-sum algorithm.
\newblock In {\em AAMAS}, 2008.

\bibitem[\protect\citeauthoryear{Farinelli \bgroup \em et al.\egroup
  }{2014}]{farinelli2014agent}
A.~Farinelli, A.~Rogers, and N.R. Jennings.
\newblock Agent-based decentralised coordination for sensor networks using the
  max-sum algorithm.
\newblock {\em Autonomous agents and multi-agent systems}, 28:337--380, 2014.

\bibitem[\protect\citeauthoryear{Fioretto \bgroup \em et al.\egroup
  }{2017}]{fioretto2017distributed}
F.~Fioretto, W.~Yeoh, E.~Pontelli, Y.~Ma, and S.J. Ranade.
\newblock A distributed constraint optimization ({DCOP}) approach to the
  economic dispatch with demand response.
\newblock In {\em AAMAS}, 2017.

\bibitem[\protect\citeauthoryear{Fitzpatrick and
  Meetrens}{2003}]{fitzpatrick2003distributed}
S.~Fitzpatrick and L.~Meetrens.
\newblock Distributed sensor networks a multiagent perspective, chapter
  distributed coordination through anarchic optimization, 2003.

\bibitem[\protect\citeauthoryear{Fransman \bgroup \em et al.\egroup
  }{2020}]{fransman2020distributed}
J.~Fransman, J.~Sijs, H.~Dol, E.~Theunissen, and B.~Schutter.
\newblock Distributed bayesian: a continuous distributed constraint
  optimization problem solver.
\newblock {\em arXiv preprint arXiv:2002.03252}, 2020.

\bibitem[\protect\citeauthoryear{Hoang \bgroup \em et al.\egroup
  }{2019}]{hoang2019new}
K.D. Hoang, W.~Yeoh, M.~Yokoo, and Z.~Rabinovich.
\newblock New algorithms for functional distributed constraint optimization
  problems.
\newblock {\em arXiv preprint arXiv:1905.13275}, 2019.

\bibitem[\protect\citeauthoryear{Hsin and Liu}{2004}]{hsin2004network}
C.n Hsin and M.~Liu.
\newblock Network coverage using low duty-cycled sensors: random \& coordinated
  sleep algorithms.
\newblock In {\em IPSN}, 2004.

\bibitem[\protect\citeauthoryear{Khan \bgroup \em et al.\egroup }{2018a}]{gene}
M.M. Khan, L.~Tran-Thanh, and N.R. Jennings.
\newblock A generic domain pruning technique for gdl-based dcop algorithms in
  cooperative multi-agent systems.
\newblock In {\em AAMAS}, 2018.

\bibitem[\protect\citeauthoryear{Khan \bgroup \em et al.\egroup
  }{2018b}]{nodeto}
M.M. Khan, L.~Tran-Thanh, W.~Yeoh, and N.R. Jennings.
\newblock A near-optimal node-to-agent mapping heuristic for {GDL}-based dcop
  algorithms in multi-agent systems.
\newblock In {\em AAMAS}, 2018.

\bibitem[\protect\citeauthoryear{Kirkpatrick \bgroup \em et al.\egroup
  }{1983}]{Kirkpatrick1983OptimizationBS}
S.~Kirkpatrick, C.D. Gelatt, and M.P. Vecchi.
\newblock Optimization by simulated annealing.
\newblock {\em Science}, 220 4598:671--80, 1983.

\bibitem[\protect\citeauthoryear{Kroese \bgroup \em et al.\egroup
  }{2011}]{Kroese2011HandbookOM}
P.D. Kroese, T.~Taimre, and Z.I. Botev.
\newblock Handbook of monte carlo methods.
\newblock 2011.

\bibitem[\protect\citeauthoryear{Maheswaran \bgroup \em et al.\egroup
  }{2004a}]{Maheswaran2004Distributed}
R.T. Maheswaran, J.P. Pearce, and M.~Tambe.
\newblock Distributed algorithms for dcop: A graphical-game-based approach.
\newblock In {\em ISCA PDCS}, 2004.

\bibitem[\protect\citeauthoryear{Maheswaran \bgroup \em et al.\egroup
  }{2004b}]{Maheswaran2004TakingDT}
R.T. Maheswaran, M.~Tambe, E~Bowring, J.P. Pearce, and P.~Varakantham.
\newblock Taking {DCOP} to the real world: Efficient complete solutions for
  distributed multi-event scheduling.
\newblock In {\em AAMAS}, 2004.

\bibitem[\protect\citeauthoryear{Mahmud \bgroup \em et al.\egroup
  }{2020}]{Mahmud2020}
S.~Mahmud, M.~Choudhury, M.M. Khan, L.~Tran-Thanh, and N.R. Jennings.
\newblock {AED}: An anytime evolutionary dcop algorithm.
\newblock In {\em AAMAS}, 2020.

\bibitem[\protect\citeauthoryear{Modi \bgroup \em et al.\egroup
  }{2005}]{modi2005adopt}
P.~J. Modi, W.~Shen, M.~Tambe, and M.~Yokoo.
\newblock Adopt: Asynchronous distributed constraint optimization with quality
  guarantees.
\newblock {\em Artificial Intelligence}, 161:149--180, 2005.

\bibitem[\protect\citeauthoryear{Okamoto \bgroup \em et al.\egroup
  }{2016}]{okamoto2016distributed}
S.~Okamoto, R.~Zivan, and A.~Nahon.
\newblock Distributed breakout: Beyond satisfaction.
\newblock In {\em IJCAI}, pages 447--453, 2016.

\bibitem[\protect\citeauthoryear{Petcu and Faltings}{2005}]{Petcu2005ASM}
A.~Petcu and B.~Faltings.
\newblock A scalable method for multiagent constraint optimization.
\newblock In {\em IJCAI}, 2005.

\bibitem[\protect\citeauthoryear{Stranders \bgroup \em et al.\egroup
  }{2009}]{stranders2009decentralised}
R.~Stranders, A.~Farinelli, A.~Rogers, and N.R. Jennings.
\newblock Decentralised coordination of continuously valued control parameters
  using the max-sum algorithm.
\newblock In {\em AAMAS}, 2009.

\bibitem[\protect\citeauthoryear{Stranders}{2010}]{Stranders2010DecentralisedCO}
R.~Stranders.
\newblock Decentralised coordination of information gathering agents.
\newblock 2010.

\bibitem[\protect\citeauthoryear{Thien \bgroup \em et al.\egroup
  }{2019}]{dgibbs}
N.~Thien, W.~Yeoh, H.~Lau, and R.~Zivan.
\newblock Distributed gibbs: A linear-space sampling-based dcop algorithm.
\newblock {\em Journal of Artificial Intelligence Research}, 64:705--748, 2019.

\bibitem[\protect\citeauthoryear{Voice \bgroup \em et al.\egroup
  }{2010}]{voice2010hybrid}
T.~Voice, R.~Stranders, A.~Rogers, and N.R. Jennings.
\newblock A hybrid continuous max-sum algorithm for decentralised coordination.
\newblock In {\em ECAI}, 2010.

\bibitem[\protect\citeauthoryear{Yokoo \bgroup \em et al.\egroup
  }{1998}]{yokoo1998distributed}
M.~Yokoo, E.H. Durfee, T.~Ishida, and K.~Kuwabara.
\newblock The distributed constraint satisfaction problem: Formalization and
  algorithms.
\newblock {\em IEEE Transactions on knowledge and data engineering},
  10:673--685, 1998.

\bibitem[\protect\citeauthoryear{Zhang \bgroup \em et al.\egroup
  }{2005}]{zhang2005distributed}
W.~Zhang, W.~Wang, Z.~Xing, and L.~Wittenburg.
\newblock Distributed stochastic search and distributed breakout: properties,
  comparison and applications to constraint optimization problems in sensor
  networks.
\newblock {\em Artificial Intelligence}, 161:55--87, 2005.

\bibitem[\protect\citeauthoryear{Zivan and Peled}{2012}]{zivan2012max}
R.~Zivan and H.~Peled.
\newblock Max/min-sum distributed constraint optimization through value
  propagation on an alternating dag.
\newblock In {\em AAMAS}, 2012.

\bibitem[\protect\citeauthoryear{Zivan \bgroup \em et al.\egroup
  }{2014}]{zivan2014explorative}
R.~Zivan, S.~Okamoto, and H.~Peled.
\newblock Explorative anytime local search for distributed constraint
  optimization.
\newblock {\em Artificial Intelligence}, 212:1--26, 2014.

\end{thebibliography}

\end{document}